\documentstyle[rawfonts,budapest,pslatex]{article}  
\frompage{000} \topage{000}                                              
\def\rightmark{Heavy Quark Production} 
\def\leftmark{R. Vogt}
 
\title{Heavy Quark Production in Heavy Ion Colliders} 
\authors{
{R. Vogt}\\[2.812mm]
{\normalsize
Physics Department, University of California, Davis, CA 95616 \\
and\\
Nuclear Science Division,
Lawrence Berkeley National Laboratory \\ 
University of California, Berkeley, California 94720, USA\\[0.2ex] 
}}
 
\abstract{We describe updated calculations of $Q \overline Q$
production in $pp$ and $\pi^- p$ interactions.  We compare these results
to total cross section data and discuss how the baseline 
cross sections extrapolate to
heavy ion collider energies. We touch upon the differences between leading
and next-to-leading order heavy quark production.  Finally, we discuss the 
implications of our calculations for quarkonium production.  Our discussion 
here focuses on bottom quarks.}

 
\begin{document}
 
\maketitle
\setcounter{page}{1}

\section{Introduction}\label{intro}

Heavy quark production is of great relevance for high energy nuclear 
collisions.  The large heavy quark masses means that their production can be 
calculated in perturbative QCD.  They are produced in the initial 
nucleon-nucleon collisions and, as such, provide valuable information about
the early state of the system.  It is therefore important to make systematic
studies to obtain the fullest possible information.  The baseline rates in 
$pp$ interactions are essential to determine the expected total cross sections
and the unmodified quark distributions.

There are many effects which can be studied through systematic heavy quark 
measurements.  Heavy quark
decays are expected to dominate the lepton pair continuum from the $J/\psi(c
\overline c)$ and $\Upsilon(b \overline b)$ up to the mass of the $Z^0$ 
\cite{gmrv,losn,cmsdoc}.
Thus the Drell-Yan yield and any thermal dilepton production will essentially
be hidden by the heavy quark decay contributions \cite{gmrv}.  The shape of the
charm and bottom contributions to this continuum could be significantly altered
by heavy quark energy loss \cite{losn,linv}. Since $B$ mesons decay to 
$J/\psi$, the secondary $J/\psi$ spectrum could be modified by energy loss 
effects on the primary $B$ \cite{losn}.  If the heavy quark energy
loss is large, it may be possible to extract a thermal dilepton yield
if it cannot be determined by other means \cite{gkp}.
Heavy quark production in a quark-gluon plasma has also been predicted
\cite{thermc}.  This additional yield can only be determined if the relative
$AA/pp$ rate can be accurately measured.  Finally, the $c \overline c$ yield 
is a useful reference for $J/\psi$ production since a
$\langle J/\psi \rangle/\langle c \overline c \rangle$ enhancement has been 
predicted due to $c \overline c$ recombination \cite{pbmjs,goren,grrapp,thews}.
This recombination requires more than one $c \overline c$ pair produced in an
event, easily reached in heavy ion collisions at RHIC and LHC.

An update to the $pp$ baseline is relevant now because the gluon distributions
have changed significantly since the calculations of Ref.~\cite{hpc}.  The
rise at low momentum fractions, $x$, is not as large as previously expected,
significantly lowering the $c \overline c$ cross sections in particular.  Less
attention has been paid to the $b \overline b$ cross sections, perhaps due to
a lack of $pp$ data.  However, since a new measurement has recently become 
available \cite{herab}, we will try to make up for this shortcoming here by 
concentrating on bottom production.

\section{$Q \overline Q$ Total Cross Sections}

At leading order (LO) heavy quarks are produced by $gg$ fusion and
$q \overline q$ annihilation while at next-to-leading order (NLO) $qg$ and
$\overline q g$ scattering is also included.  To any order, the partonic 
cross section may be expressed in terms of dimensionless scaling functions
$f^{(k,l)}_{ij}$ that depend only on the variable $\eta$ \cite{KLMV},
\begin{eqnarray}
\label{scalingfunctions}
\hat \sigma_{ij}(\hat s,m_Q^2,\mu^2) = \frac{\alpha^2_s(\mu)}{m^2}
\sum\limits_{k=0}^{\infty} \,\, \left( 4 \pi \alpha_s(\mu) \right)^k
\sum\limits_{l=0}^k \,\, f^{(k,l)}_{ij}(\eta) \,\,
\ln^l\left(\frac{\mu^2}{m_Q^2}\right) \, , 
\end{eqnarray} 
where $\hat s$ is the partonic center of mass energy squared, 
$m_Q$ is the heavy quark mass,
$\mu$ is the scale and $\eta = \hat s/4 m_Q^2 - 1$.  
The cross section is calculated as an expansion in powers of $\alpha_s$
with $k=0$ corresponding to the Born cross section at order ${\cal
O}(\alpha_s^2)$.  The first correction, $k=1$, corresponds to the NLO cross
section at ${\cal O}(\alpha_s^3)$.  It is only at this order and above that
the dependence on renormalization scale, $\mu_R$, enters the calculation
since when $k=1$
and $l=1$, the logarithm $\ln(\mu^2/m_Q^2)$ 
appears.  The dependence on the factorization scale, $\mu_F$, the argument of
$\alpha_s$, appears already at LO.  We assume that $\mu_R = \mu_F = \mu$.  
The next-to-next-to-leading
order (NNLO) corrections to next-to-next-to-leading logarithm
have been calculated near threshold \cite{KLMV} but
the complete calculation only exists to NLO.

The total hadronic cross section is obtained by convoluting the total partonic
cross section with the parton distribution functions (PDFs)
of the initial hadrons,
\begin{eqnarray}
\label{totalhadroncrs}
\sigma_{pp}(s,m_Q^2) = \sum_{i,j = q,{\overline q},g} 
\int_{\frac{4m_Q^2}{s}}^{1} \frac{d\tau}{\tau}\, \delta(x_1 x_2 - \tau) \,
F_i^p(x_1,\mu^2) F_j^p(x_2,\mu^2) \, 
\hat \sigma_{ij}(\tau ,m_Q^2,\mu^2)\, , 
\end{eqnarray}
where the sums $i$ and $j$ are over all massless partons and
$x_1$ and $x_2$ are fractional momenta.
The PDFs, denoted by $F_i^p$, are evaluated at
scale $\mu$.  All our calculations are fully NLO, applying NLO parton
distribution functions and the two-loop evaluation of
$\alpha_s$ to both the ${\cal
O}(\alpha_s^2)$ and ${\cal O}(\alpha_s^3)$ contributions, as is typically done
\cite{KLMV,MNR}.

\begin{figure}[htbp]
\setlength{\epsfxsize=0.75\textwidth}
\setlength{\epsfysize=0.5\textheight}
\centerline{\epsffile{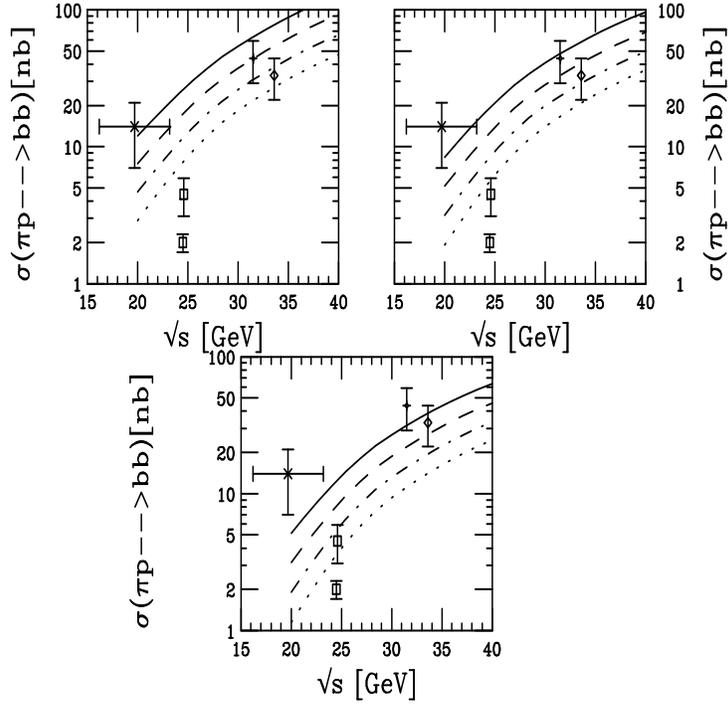}}
\caption[]{Total $b \overline b$ cross sections in $\pi^-p$ interactions at
fixed-target energies as a function of the bottom quark mass.  
All calculations are fully 
NLO using the MRST HO (central gluon) parton densities.  From the upper left,
the plots show results with the renormalization and factorization scales equal 
to $m_b/2$, $m_b$ and $2m_b$ respectively.  In each plot, from top
to bottom the curves are $m_b = 4.25$, 4.5, 4.75, and 5 GeV.  The $b \overline
b$ data can be found in Ref.~\protect\cite{hpc}.
}
\label{pipbbvmb}
\end{figure}

To obtain the $pp$ cross sections at the RHIC and LHC proton and ion energies, 
we first compare the NLO
cross sections to the available $c \overline c$ and $b \overline b$ production
data by varying the mass, $m_Q$, and scale, $\mu$, to obtain the `best'
agreement with the data for several combinations of $m_Q$, $\mu$, and PDF.
These `best' fits aim to describe the available data without a scaling factor,
the experimental `$K$' factor.
We use the recent MRST HO central gluon \cite{mrst}, CTEQ 5M \cite{cteq5}, and
GRV 98 HO \cite{grv98} distributions.  The results for the $b \overline b$
cross section in $\pi^- p$ interactions, where there are the most data, 
are shown in Fig.~\ref{pipbbvmb} for the MRST HO densities with the SMRS
pion densities \cite{SMRS}.  
We calculate the NLO cross sections for $4.25 \leq m_b \leq 5$ GeV with scale
choices clockwise from the upper left: $\mu = m_b/2$, $m_b$, and $2m_b$.  
The cross sections decrease as $\mu$ increases because $\alpha_s(m_b/2) >
\alpha_s(m_b) > \alpha_s(2m_b)$ by the running of $\alpha_s$.  
Evolution of the PDFs
with $\mu$ tends to go in the opposite direction {\it e.g.}\ the sea quark
and gluon distributions rise as $x$ decreases while $\mu$ increases.  
At higher scales the two
effects tend to compensate and reduce the scale dependence but the charm quark
mass is not large enough for this to occur. The results for bottom quarks are
somewhat better since the mass is larger. 

\begin{figure}[tbh]
\setlength{\epsfxsize=0.9\textwidth}
\setlength{\epsfysize=0.3\textheight}
\centerline{\epsffile{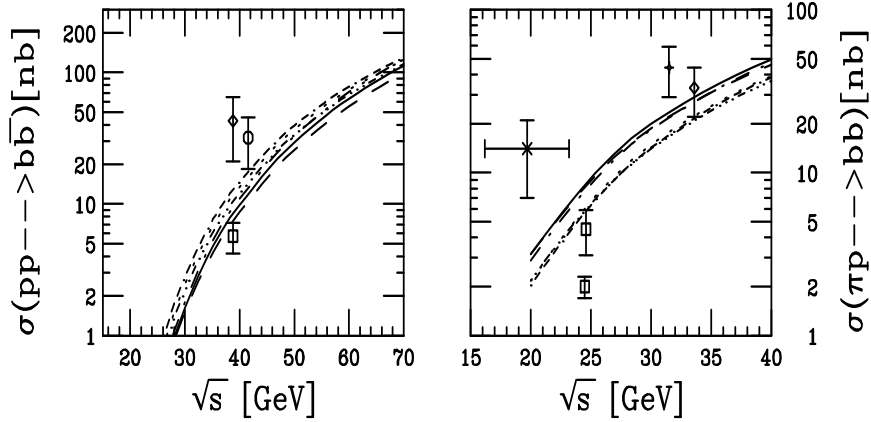}}
\caption[]{Total $b \overline b$ cross sections in $pp$ and $\pi^- p$
interactions compared to data.  All calculations are fully 
NLO.  The curves are: MRST HO (central gluon) with $m_b = \mu = 4.75$ GeV 
(solid), $m_b = \mu/2 = 4.5$ GeV (dashed), and $m_b = 2\mu = 5$ GeV 
(dot-dashed); as well as the GRV 98 HO distributions with 
$m_b = \mu = 4.75$ GeV (dotted), $m_b = \mu/2 = 4.5$ GeV (dot-dot-dot-dashed), 
and $m_b = 2\mu = 5$ GeV (dot-dash-dash-dashed). 
The $pp$ data are from the E789 (square) \protect\cite{E789}, E771
(diamond) \protect\cite{E771} and preliminary HERA-B (circle) 
\protect\cite{herab} collaborations.}
\label{pppipbb}
\end{figure}

We find reasonable agreement with all three PDFs for $m_b = \mu = 4.75$ GeV, 
$m_b = \mu/2 = 4.5$ GeV (dashed), and $m_b = 2\mu = 5$ GeV, shown in the right
hand side of Fig.~\ref{pppipbb}.  The MRST results cluster together and
lie a bit higher than the GRV 98 results for $\pi^- p$ production while the 
opposite is true for $pp$ production.  The three data points from $pp 
\rightarrow b \overline b$ measurements lie somewhat closer together although
the E789 point (the square on the left hand side of Fig.~\ref{pppipbb}) 
lies considerably below the other two points.  In
fact, it agrees best with the NLO calculations using a `standard'
$b$ quark mass of 4.75 GeV.

The somewhat larger spread in the $\pi^- p$ calculations may be
because the $\pi^-$ PDFs are not very well known.  The last evaluations, SMRS
\cite{SMRS}, Owens-$\pi$ \cite{Owenspi}, and GRV-$\pi$ \cite{GRVpi} were 10-15
years ago and do not reflect any of the latest information on the low $x$
behavior of the proton PDFs, {\it e.g.}\ the distributions are all flat as $x
\rightarrow 0$ with no low $x$ rise.  The GRV distributions are based on 
dynamical parton densities with low initial scale which generate larger 
densities at low $x$ while depleting them at high $x$, reducing the 
$b \overline b$ cross sections at low $\sqrt{s}$ relative to the flat 
distributions of SMRS, see Fig.~\ref{pppipbb}.
These pion evaluations also depend on the
behavior of the proton PDFs used in the original fit, including the value of
$\Lambda_{\rm QCD}$.  Thus the pion and proton PDFs are generally incompatible.
The
typical $x$ values of $b \overline b$ production are large, $0.14 \leq x = 
2\mu/\sqrt{s} \leq 0.9$.  At the fixed target energies of $b \overline b$
production, $q \overline q$ annihilation dominates while $gg$ fusion is still
most important for $c \overline c$ production \cite{smithrv}.  The
valence-valence $\overline u_\pi u_p$ contribution dominates since the
valence distributions are greatest at large $x$. 

For charm production,
the `best' agreement with $\mu =
m_c$ is for $m_c = 1.4$ GeV and $m_c = 1.2$ GeV is the best choice for $\mu =
2m_c$ for the MRST HO and CTEQ 5M distributions.  The best agreement with GRV
98 HO is $\mu = m_c = 1.3$ GeV while the results with $\mu = 2m_c$ lies below
the data for all $m_c$.  All five results agree very well with each other for
$pp \rightarrow c \overline c$.  
There is also more of a spread in the $\pi^- p \rightarrow c
\overline c$ results. The $\pi^- p \rightarrow c \overline c$ 
cross sections are a bit lower than the data compared
to the $pp$ cross sections, suggesting that lighter quark masses would tend to
be favored.  The reason is because the low $x$ rise in the proton
PDFs depletes the gluon density for $x > 0.02$ relative to a constant at $x
\rightarrow 0$ for $\mu = \mu_0$ the initial scale of the PDF.  The $\pi^- p$
data are also in a relatively large $x$ region, $0.1 \leq x = 2\mu/\sqrt{s} 
\leq 0.3$, where this difference is important.  For the $c \overline c$
results, see Ref.~\cite{RVww}.

Before calculating the $Q \overline Q$
cross sections at nuclear colliders, some comments need
to be made about the validity of the procedure.  
Since the $c \overline c$ 
calculations can only be made to agree with the data when somewhat lower than
average quark masses are used and even the $pp \rightarrow b \overline b$ data
suggest $m_b$ should be smaller, it is reasonable to expect that higher order
corrections beyond NLO could be large.  Indeed, the preliminary HERA-B 
cross section agrees with the NNLO-NNLL cross section in 
Ref.~\cite{KLMV}, suggesting that the next order correction could be nearly a
factor of two.  Thus the NNLO correction could be nearly as large as the NLO
cross section.

Unfortunately, the NNLO-NNLL calculation is not a full result
and is valid only near threshold.  The $p \overline p$ data at higher energies,
while not total cross sections, also show a large discrepancy between the
perturbative NLO result and the data, nearly a factor of three \cite{cdf}.  
This difference could be accounted for using unintegrated parton densities
\cite{unint} although these unintegrated distributions vary widely at this
point.  The problem is then how to connect the regimes where 
near-threshold corrections are applicable and where high-energy, small $x$ 
physics dominates.  The problem is increased for charm where, even at low
energies, we are far away from threshold.  

Our method is perhaps the
easiest thing to try--using NLO only and ignoring higher-order corrections to
fit the data.  This is not difficult for $c \overline c$ because 
the data are extensive.  However, there
is less $b \overline b$ data to go on.  The $\pi^- p \rightarrow c \overline
c$ data tends to favor lighter masses.  It is difficult to say if the same is
true for $b \overline b$.  A value of $m_b = 4.75$ GeV, which underpredicts
the Tevatron results compared to NLO cross sections \cite{cdf}, agrees 
reasonably with the average of the $\pi^- p$ data.  However, for the HERA-B
measurement to be compatible with a NLO evaluation, the $b$ quark mass would
have to be reduced to 4.25 GeV, a value which might be more compatible with
the Tevatron results.  Therefore, if the NNLO cross section could be calculated
fully, the results would likely be more compatible with a larger quark mass.
A quantitative statement is not possible at this time.

If we then assume that the NNLO and higher orders to not have 
a substantially different energy dependence than the LO and NLO results, 
then we will be in the right ballpark at collider energies.  If the LO and 
NLO matrix elements are both evaluated with NLO
PDFs and the two-loop $\alpha_s$, the theoretical $K$ factors have 
a relatively weak $\sqrt{s}$ dependence, $\leq 50$\% for $20 \, {\rm
GeV} \leq \sqrt{s} \leq 14 \, {\rm TeV}$.  The produced heavy
quark distributions might be 
slightly affected since the shapes are somewhat sensitive to the quark mass 
but charm is far enough above threshold at ion colliders for the effect to be
small.  A somewhat larger effect might be expected for bottom.

\begin{figure}[tbh]
\setlength{\epsfxsize=0.65\textwidth}
\setlength{\epsfysize=0.3\textheight}
\centerline{\epsffile{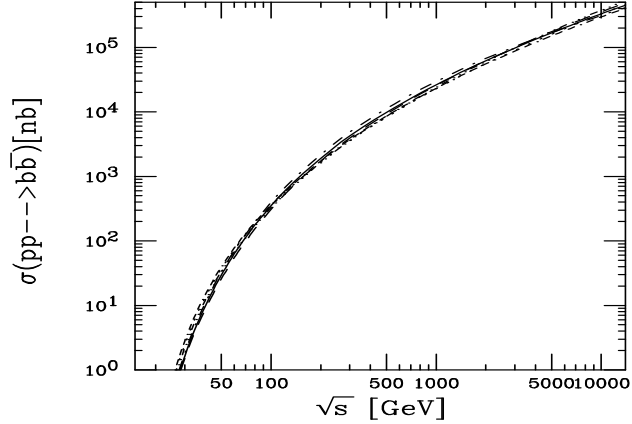}}
\caption[]{Total $b \overline b$ cross sections in $pp$
interactions up to 14 TeV.   All calculations are fully 
NLO.  The curves are the same as in Fig.~\ref{pppipbb}.}
\label{ppbblhc}
\end{figure}

We now extrapolate our results to RHIC and LHC energies.  The 
$b \overline b$ cross section is shown in Fig.~\ref{ppbblhc}.  
The spread in the $b \overline b$ cross sections is
small, $\sim 20-30$\% at the ion collider energies.  The $c \overline c$
cross sections, on the other hand, differ by a factor of two at 5.5 TeV.

Defining the theoretical $K$ factor as NLO/LO, where both the numerator and
denominator are calculated with NLO PDFs and the two-loop $\alpha_s$, is not
the only way to proceed.  It is, however, the most common because, in general,
one wants to determine the size of the corrections.  In principle, it is most
correct to  use LO PDFs and the one-loop $\alpha_s$ with the LO matrix elements
and NLO PDFs and the two-loop $\alpha_s$ only with the ${\cal O}(\alpha_s^3)$
corrections.  Since the QCD
scale, $\Lambda$, is tuned to fit all available data in global analyses of
parton distributions, whether at LO or NLO, the number of loops in the
evaluation of $\alpha_s$ is important.  When a fully LO 
calculation of the charm cross 
section is done with a one loop evaluation of
$\alpha_s$ in the global analyses of the parton densities, as with the MRST LO
set, the LO cross sections increase by $\sim 60$\%.
This change is almost solely
due to the difference between the one and two loop evaluations of
$\alpha_s$.  
The MRST LO $\Lambda$ is 0.204 GeV when $n_f=3$, leading
to $\alpha_s^{\rm 1-loop} = 0.287$ and $\alpha_s^{\rm 2-loops} = 0.220$
for $\mu = 2m_c$.
The $\Lambda$ associated with the MRST HO set is larger, $\Lambda = 0.353$ GeV
when $n_f = 3$, corresponding to $\alpha_s^{\rm 1-loop} = 0.364$ and 
$\alpha_s^{\rm 2-loops} = 0.263$ at the same scale.  
The shapes of the LO and 
NLO PDFs are also somewhat different which can affect results in different
regions.

A third way of defining the theoretical $K$ factor is also possible.  One
could compare a fully NLO calculation, where the LO and NLO matrix elements
are calculated with $\alpha_s$ at two loops and a NLO PDF, with a LO
calculation where the LO matrix elements are calculated with $\alpha_s$ at
one loop and a LO PDF.  This procedure is possibly most favored {\it e.g.}\
in event generators where most processes are calculated with LO matrix elements
only.  The magnitude of the theoretical
$K$ factor depends on the chosen definition.
Work is in progress to compare all three ways of evaluating the $K$
factor for heavy quark production \cite{rvnewk}.

Note that while the total cross section 
predicts the yield of heavy quark production over all phase space,
it cannot provide much useful information on nuclear effects such a
$p_T$-broadening and shadowing.  
Any broadening will not affect the total cross
section but will have a strong influence on the $p_T$ distributions.
Shadowing may reduce or enhance the nuclear
cross section relative to that of the proton 
but the effect may be more apparent in
some regions of phase space than others.
To obtain more information on nuclear effects, it is thus
necessary to turn to distributions.  In addition, a real detector does not 
cover all phase space.  The differential distributions can be tuned to a 
detector acceptance.  For full details on the shadowing and broadening effects
on the single quark and $Q \overline Q$ pair distributions, see 
Ref.~\cite{RVhpc}. It turns out that the $p_T$ distributions are more
strongly influenced by any broadening effect than nuclear shadowing.  Likewise,
nuclear shadowing can best be studied combining information on rapidity and 
pair invariant mass distributions.  The effect of shadowing on the dilepton 
continuum has been extensively studied in Ref.~\cite{EKV}.

\section{Relevance for Quarkonium}

To better understand quarkonium in nuclear collisions, it is 
necessary to have a good estimate of the expected $pp$ yields.  
However, there are still a number of 
uncertainties in quarkonium production in $pp$ interactions.  
Two approaches have been used to describe 
quarkonium
production phenomenologically---the color evaporation model (CEM) \cite{bkp}
and nonrelativistic QCD (NRQCD) \cite{nrqcd}.  

In the CEM, the $Q \overline Q$ pair 
neutralizes its color by 
interaction with the collision-induced color field---``color evaporation".
The $Q$ and the $\overline Q$ either combine with light
quarks to produce heavy-flavored hadrons or bind with each other 
in a quarkonium state.  The additional energy needed to produce
heavy-flavored hadrons is obtained nonperturbatively from the
color field in the interaction region.
The yield of all quarkonium states
may be only a small fraction of the total $Q\overline 
Q$ cross section below the heavy hadron threshold, $2m_H$.
The $Q \overline Q$ cross sections we have obtained through our `by-eye' fits
of the mass and scale parameters have implications for quarkonium if the CEM
is used to calculate the production.  This is because different quark masses
will result in more or less of the cross section below the $2m_H$ threshold.
Since we have concentrated on $b \overline b$ production in these proceedings,
we show our results for $\Upsilon$ production where there is the biggest 
variation in mass and scale: $m_b = 4.5$ GeV is $0.85 \, m_B$ while $m_b = 5$
GeV is $0.95 \, m_B$, much closer to the $B \overline B$ threshold.  Since
the NRQCD cross section is obtained independently of the parameters used to 
calculate heavy quark production, we do not discuss it here.

\begin{figure}[htb] 
\setlength{\epsfxsize=0.65\textwidth}
\setlength{\epsfysize=0.4\textheight}
\centerline{\epsffile{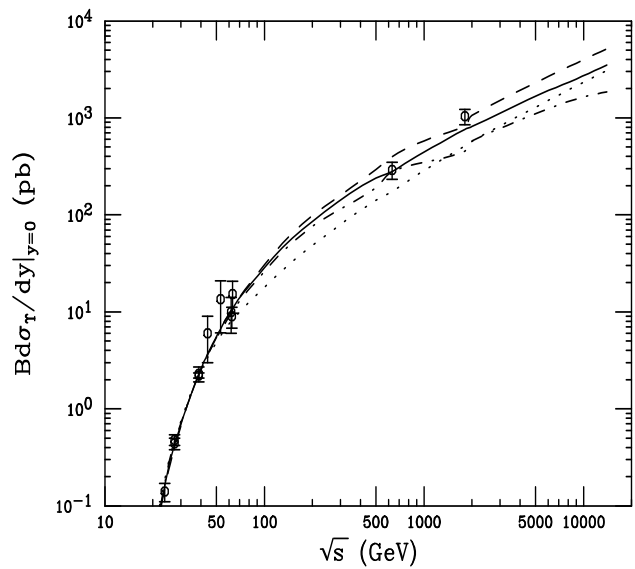}}
\caption[]{Inclusive $\Upsilon$ production data, combined from all three $S$
states, and compared to NLO CEM calculations.  The
solid curve employs the MRST HO distributions with $m_b = \mu = 4.75$ GeV,
the dashed, $m_b = \mu/2 = 4.5$ GeV, the dot-dashed, 
$m_b = 2\mu = 5$ GeV, and the dotted, GRV 98 HO with $m_b = \mu = 4.75$ GeV.}
\label{upsplot}
\end{figure}

At leading order, the production cross section of quarkonium state $C$ is
\begin{eqnarray}
\sigma_C^{\rm CEM} = F_C \sum_{i,j} \int_{4m_Q^2}^{4m_H^2} d\hat s \int dx_1 
dx_2~f_{i/p}(x_1)~f_{j/p}(x_2)~ \hat\sigma_{ij}(\hat s)~\delta(\hat 
s-x_1x_2s)\, \, , \label{sigtil}
\end{eqnarray} 
where $ij = q \overline q$ or $gg$ and $\hat\sigma_{ij}(\hat s)$ is the
$ij\rightarrow Q\overline Q$ subprocess cross section.
Hadronization is assumed not to affect the kinematics of the parent
$Q \overline Q$ pair so that only a
single universal factor, $F_C$, is necessary for the cross section of
each state.
The factor $F_C$ depends on the heavy quark mass, $m_Q$,
the scale $\mu$ in the strong coupling constant $\alpha_s$ and the
parton densities.  The factor $F_C$ must be
constant for the model to have any predictive power.
The CEM was taken to NLO using exclusive $Q \overline
Q$ hadroproduction
\cite{MNR} to obtain the energy, $x_F$, and $p_T$-dependence
of quarkonium production \cite{Gavai,SchulV}.  In the CEM, 
$gg \rightarrow g (g^* \rightarrow Q \overline Q)$, incorporated at NLO,
is similar to models of $g^* \rightarrow \Upsilon$
fragmentation \cite{frag}.  By including
this splitting, the CEM provides a good
description of the quarkonium $p_T$ distributions at the Tevatron.  The CEM
cross sections were calculated with the MRST HO \cite{mrst} parton densities
using the same values as the NLO evaluations of the heavy quark 
cross sections shown here.
The values of $F_C$ for charmonium and bottomonium have been calculated from
a fit to the $J/\psi$ and $\Upsilon$ data combined with relative cross sections
and branching ratios, see Refs.~\cite{gunv,dpsseq} for details.  The combined
$\Upsilon$, $\Upsilon'$ and $\Upsilon''$ cross sections to muon pairs are 
compared to our CEM calculations in Fig.~\ref{upsplot}.  We find that
the direct $\Upsilon(1S)$ production cross section is $\approx 3.4$ nb at
$\sqrt{s} = 200$ GeV and varies between 126-259 nb
at 5.5 TeV.

\section*{Acknowledgements}

I would like to thank the HERA-B collaboration for keeping me informed about
their exciting new data.
This work was supported by the Director,
Office of Energy Research,
Office of High Energy and Nuclear Physics,
Nuclear Physics Division of the U.S.\ Department of Energy
under Contract No.\ DE-AC03-76SF00098.

\vfill\eject
\end{document}